\documentclass[
twocolumn,
showpacs,preprintnumbers,prd,amsmath,amssymb,aps,floats]{revtex4}
\usepackage{graphicx}
\usepackage{dcolumn}
\usepackage{bm}
\def\nnd{\end{document}}
\def\dgr{\dagger}

\def\nnb{\nonumber}
\def\de{\delta}

\def\be{\begin{equation}}
\def\ee{\end{equation}}
\def\mn{\mu\nu}

\newcommand{\bea}{\begin{eqnarray}}
\newcommand{\eea}{\end{eqnarray}}
\newcommand{\bwt}{\begin{widetext}}
\newcommand{\ewt}{\end{widetext}}

\def\sbra#1{\Big   ( #1  \Big   ) }
\def\mbra#1{\bigg  [ #1  \bigg  ] }
\def\bbra#1{\Bigg \{ #1  \Bigg \} }

\def\cma{\,,}
\def\hA{\widehat A}

\def\hWp{\widehat W^{+}}
\def\bWp{W^{+}}
\def\hWm{\widehat W^{-}}
\def\bWm{W^{-}}
\def\hZ{\widehat Z}
\def\bZ{Z}
\def\crho{\overline \rho}
\def\qrho{\widehat \rho}
\def\qh{\widehat h}

\def\wsep{ \nnb \\ &&}
\def\nwsep{}

\def\eed{\end{document}}
\def\ch{ h}
\def\lx{{\stackrel{\leftharpoonup}{X}}}
\def\rx{{\stackrel{\rightharpoonup}{X}}}
\def\al{\alpha}

\def\al{\alpha}
\def\be{\beta}

\def\al{{\alpha}}

\def\bbbkf#1{\bigg \{ #1 \bigg \}}

\makeatother

\begin{document}
\preprint{ KEK-TH-1041 }

\title{A correct renormalization procedure for the electroweak chiral Lagrangian}
\author{Qi-Shu Yan \,\, }
\affiliation{Theory Group, KEK,  Tsukuba, 	 305-0801, Japan.}



\begin{abstract}
We perform a systematic one-loop renormalization 
to the standard model with a nonlinearly realized 
Higgs boson field and the electroweak chiral Lagrangian up to $O(p^2)$ order.
We find even in the nonlinear form, the Higgs model of the standard model
is renormalizable in our procedure. We also 
examine $\beta$ functions of gauge couplings for the
standard model and $\beta$ functions for anomalous couplings in 
the electroweak chiral Lagrangian up to $O(p^2)$ order,
and find our results agree with all well-known results.
Based on these observations, we conclude that our calculation procedure could
be a correct one to systematically construct the divergences of the electroweak
chiral Lagrangian when $O(p^4)$ operators are included.
\end{abstract}
\maketitle


\section{Introduction}
How to perform the systematic renormalization of the electroweak chiral 
Lagrangian (EWCL) \cite{Appelquist:1980vg,Appelquist:1980ae,Longhitano:1980iz,Longhitano:1980tm,Appelquist:1980ix,Appelquist:1993ka} including
dimensionless anomalous couplings is 
a long-time unsolved puzzle \cite{DeRujula:1991se,Hagiwara:1992eh,Hagiwara:1993ck,Burgess:1992va,Burgess:1992gx,Hernandez:1993pp,Burgess:1993qk,Dawson:1994fa,vanderBij:1997ec}. 
In order to obtain the correct answer, 
the correct renormalization procedure is very crucial. 
Otherwise, it seems confusing that starting with one unique theory, the EWCL, we can
have different answers, as shown in literatures.

While according to
statements of the effective field theory method \cite{Weinberg:1978kz,Weinberg:1995mt,georgi}, even 
with the nonrenormalizable theory, in principle, the radiative correction should be 
well-defined by using the renormalization group method. Based on the validity and great
success of the effective field theory method in B physics \cite{Buchalla:1995vs} and 
low energy hadronic physics \cite{Gasser:1983yg,Gasser:1984gg}, etc,
we believe that there exists a unique correct answer about the radiative corrections
of anomalous couplings in the EWCL.

It is well-known that if a theory is not renormalizable, in order to construct the
renormalization group equations (RGE), a correct renormalization procedure
which can reliably extract divergences is very crucial and important.
In a gauge theory, the divergences of renormalization constants are
gauge dependent, but the one-loop $\beta$ functions of the gauge couplings
are gauge independent. Therefore, even in a nonrenormalizable theory (say, the EWCL), 
we expect that we can construct gauge independent one-loop $\beta$ functions for anomalous
couplings. We are fully aware of the fact that gauge choosing is very important for
gauged nonlinear sigma models, as pointed
out by Einhorn \cite{Einhorn:1991zu,Einhorn:1991rz} 
and as we learn from the
history of renormalization of the
massive Yang-Mills theories \cite{Lee:1962vm,Appelquist:1980ix} and the standard model \cite{Veltman:1992hp}.
For example, the calculations in unitary gauge \cite{Burgess:1993qk,Dawson:1994fa} 
(a nonrenormalizable gauge \cite{Weinberg:1973ew,Veltman:1992hp}) by using the
dimensional regularization scheme with $O(p^2)$ operators can not produce the same 
divergences of those obtained in Landau gauge \cite{Longhitano:1980iz,Longhitano:1980tm,Herrero:1993nc,Herrero:1994iu,Dittmaier:1995ee}, 
and therefore have the difficulty to produce the
"screening effects" of a decoupled Higgs, as dubbed by Veltman \cite{Veltman:1976rt}.
Therefore we believe that a correct systematic renormalization procedure should 
choose a renormalizable gauge and at least pass the following criteria: 
to demonstrate the renormalizability of the SM, 
to reproduce the correct divergences of $O(p^2)$ operators, to generate the correct
$\beta$ functions for the gauge couplings, etc.

This paper is devoted to investigate our
renormalization procedure and examine whether it can pass these criteria.
Before launching on the systematic one-loop renormalization of the EWCL 
including all dimensionless anomalous couplings \cite{preparation}, we regard that 
it is a necessary and constructive step for us to first examine two
simple cases: 1) the nonlinearly realized Higgs model and 2) the
EWCL up to $O(p^2)$ operators. We will explore
two types of power counting rules in our
calculation: Weinberg's derivative power counting rule \cite{Weinberg:1978kz}
and Georgi-Manohar mass dimension power counting rule (naive dimension analysis) \cite{Manohar:1983md}.
We will also introduce the superficial divergence power counting rule which is
important for the construction of divergences.

By using the background field method \cite{DeWitt:1967ub,DeWitt:1967uc}, 
the Stueckelberg transformation \cite{Stuckelberg:1933,Grosse-Knetter:1992nn}, path integral, 
heat kernel \cite{Avramidi:2000bm,Vassilevich:2003xt} and dimensional regularization method, 
and ${\overline {MS}}$
renormalization scheme, 
we show how to perform the systematic renormalization 
at one-loop level with a nonlinearly realized Higgs 
boson and the EWCL in $O(p^2)$ operators without a Higgs boson.
New points in this paper include: 
1) Compared with calculation in the momentum space \cite{Dittmaier:1995ee}, 
some nonperturbative contributions
can be reliably extracted out in coordinate 
space for the nonlinearly realized Higgs model;
2) With the help
of the classic equation of motions of the vector and Higgs bosons 
and with a proper parameterization of the Goldstone
fields, the nonlinearly realized standard model with a Higgs boson is still 
renormalizable in one-loop level at least; this point is not trivial, there is
no reference which has shown us that the standard model with  a nonlinear
realized standard model is renormalizable.
3) Weinberg's derivative power counting rule and Georgi-Manohar's mass dimensional
analysis are compatible with each other in the nonlinearly realized SM with a Higgs boson;
4) Gauge fixing term for vector bosons
can be defined in either mass eigenstates or weak interaction eigenstates due to
the global symmetry in ghost sector. 
5) The $\beta$ functions of the gauge couplings in our calculation procedure
agree with well-known results \cite{Gross:1973id,Politzer:1973fx,'tHooft:1973us}.
6) Our calculation procedure reproduces the well-known divergences of $O(p^2)$ operators
at one-loop level \cite{Longhitano:1980iz,Longhitano:1980tm,Herrero:1993nc,Herrero:1994iu,Dittmaier:1995ee}.
7) We also show how to construct the ghost BRST transformations in the background
field method with and without a nonlinearly realized Higgs fields, 
which might be of special importance to understand the higher order  
renormalizability of the nonlinearly realized SM with a Higgs boson 
and the EWCL up to $O(p^2)$ order. 

Based on these points, we conclude that our renormalization procedure is suitable and
trustable in constructing divergences of the EWCL, even though such a theory is a 
nonrenormalizable theory in the perturbation expansion due 
to the nonlinearly realized Goldstone bosons.

As far as we know, all the techniques, methods, and concepts demonstrated in this paper
have been used in the hadronic chiral Lagrangian, up to
$O(p^4)$ order \cite{Gasser:1983yg,Gasser:1984gg}, and up to $O(p^6)$
order calculations\cite{Bijnens:1999sh,Bijnens:1999hw}, 
two loop renormalization in gravity in 4 dimension \cite{vandeVen:1991gw},
three loop $\beta$ functions in Yang-Mills theory \cite{Bornsen:2002hh},
the one-loop renormalization in the $\pi-\rho$ system \cite{Tanabashi:1993sr,Harada:2003jx},
and the gauged nonlinear $SU(2)$ sigma model 
including dimensionless anomalous couplings \cite{Yan:2002rt}.

For the sake of consistency and simplicity, we avoid the Wick rotation in the
loop integral and simply
work in the Euclidean coordinate space to extract divergences. 
We have checked our formula for extracting divergences in 
the coordinate space with those provided in \cite{Jack:1982hf,vandeVen:1991gw} and found agreement.

\section{the $SU(2) \times U(1)$ standard model with nonlinearly realized Higgs boson}
The classic Lagrangian of the standard $SU(2)\times U(1)$ gauge theory without including Fermions 
can be formulated as
\bea
\label{eq:sm}
{ \cal L} &=& - H_1 - H_2 - (D\phi)^{\dagger}\cdot(D \phi) \wsep
+\mu^2 \phi^{\dagger} \phi - {\lambda \over 4} (\phi^{\dagger} \phi)^2 \cma\\
H_1&=&{1\over 4}  W_{\mn}^a W^{a \mn}\cma\\
H_2&=&{1\over 4}  B_{\mn}   B^{\mn}\cma
\eea
where the $W$ and $B$ are the vector bosons of $SU_L(2)$ and $U_Y(1)$ gauge
groups, respectively. The $\phi$ is the Higgs field, a weak doublet.
The $\mu^2$ and $\lambda$ are two variables of the Higgs potential, which
determine the spontaneous breaking of symmetry. 
The definitions of other quantities are standard and listed as 
\bea
W_{\mn}^a&=&\partial_{\mu} W_{\nu}^a - \partial_{\nu} W_{\mu}^a +g f^{abc} W_{\mu}^b W_{\nu}^c\,,\\
B_{\mn} &=&\partial_{\mu} B_{\nu} - \partial_{\nu} B_{\mu} \cma \\
D_{\mu} \phi&=&\partial_{\mu} \phi - i g W_{\mu}^a T^a + i y_{\phi} g' B_{\mu} T^3 \phi\,,\\
\phi^{\dgr}&=&(\phi_1^*,\phi_2^*)\,,
\eea
where $T^a$ are the generators of the Lie algebra of
$SU(2)$ gauge group, and $a=1,\,2,\,3$. The
Y charge of the field $\phi$ is given as $y_{\phi}=1$.
The $g$ and $g'$ are the couplings of the corresponding
gauge interactions, respectively.

The spontaneous symmetry breaking is realized by taking 
the mass square $\mu^2$  to be positive.
The vacuum expectation value of Higgs
field is solved from the Higgs potential
as $|\langle \phi \rangle| = v/{\sqrt 2}$.
According to the Goldstone theorem, there are
three Goldstone bosons which break break $SU_L(2)\times U_Y(1)$ down to
$U_{\textrm{EM}}(1)$ symmetry. And by eating the corresponding
Goldstone bosons, the vector bosons $W$ and $Z$ obtain their masses, while
the vector bosons $A$ of the unbroken $U(1)_{em}$ gauge symmetry are still massless.

The Lagrangian given in Eq. (\ref{eq:sm})
with the Higgs mechanism can be reformulated in its nonlinear
form by changing the variable $\phi$
\bea
\label{eq:chv}
\phi={1\over \sqrt{2}} (v + h) U\,,\,
U=\exp\left ({\bf 2} {i \zeta^a T^a }\right )\,,\,
v=2 \sqrt{\mu^2 \over \lambda}\,,
\eea
where the $h$ is the Higgs scalar, $v$ is the vacuum expectation value.
The $U$ is a phase factor, and the $\zeta^a,\,a=1,\,2,\,3$ are the corresponding
phases which can be parameterized as the Goldstone bosons,
as prescribed by the Goldstone theorem. The exact form for this parameterization will
be given latter.

As we know, the change of variables in Eq. (\ref{eq:chv}) induces
a determinant factor in the functional integral ${\cal Z}$
\bea
{\cal Z} = \int {\cal D} W_{\mu}^a {\cal D} h {\cal D}\xi^b
\exp\sbra{{\cal S}'[W,h,\xi]} \\
\times \det\bbra{ \sbra{1 + {h \over v}} \delta(x-y) }\,,
\eea
and correspondingly modifies the Lagrangian density to
\bea
\label{hml2}
{\cal L} &=& - H_1 - H_2 - {(v+h)^2 \over 4} tr[D U^{\dagger} \cdot D U] \nnb\\
&& - {1\over 2} \partial h \cdot \partial h
+ {\mu^2 \over 2 } (v + h)^2 - {\lambda \over 16} (v+h)^4
\nnb \\ && + \delta^4(0) ln \left \{ 1 + {h\over v} \right \}\,.
\label{uglag}
\eea
As pointed out by several references \cite{ugauge,Grosse-Knetter:1992nn},
this determinant containing quartic divergences is
indispensable and crucial to cancel exactly the quartic
divergences brought into by the longitudinal part
of vector bosons, and is important in verifying the
renormalizability of the Higgs model in the U-gauge
and the equivalence of U-gauge to other gauges.

There are also quartic divergent terms induced by $\zeta$ as shown in \cite{Grosse-Knetter:1992nn},
here we have omitted them.

The Lagrangian ${\cal L}$ given in Eq. (\ref{uglag})
is invariant under the following local $SU(2)_L\times U(1)_Y$ chiral transformation
 \bea
\label{}
U &\rightarrow& g_L U g_R^\dagger \cma\nnb\\
W_\mu &\rightarrow& g_L^\dagger  W_\mu g_L+ i {1\over g} g_L^\dagger \partial_\mu g_L \cma \nnb\\
W_{\mu\nu}&\rightarrow& g_L^\dagger W_{\mu\nu} g_L \cma \nnb \\
B_\mu &\rightarrow& B_\mu + i {1 \over g'} g_R^\dagger \partial_\mu g_R\cma \nnb \\
B_{\mu\nu}&\rightarrow&  B_{\mu\nu}\cma\nnb \\
v+ \ch &\rightarrow & v + \ch\cma
\eea
where the gauge transformation factors $g_L$ and $g_R $ are defined as
\bea
g_L &=& \exp{\bbra{- i g {\alpha^{a}}_L T^a } } \,,\nnb\\
g_R &=& \exp{\bbra{- i g'      \beta_R T^3 } } \,.
\eea
Here $\alpha^a$ and $\beta_R$ are parameters of gauge group transformations.

In order to contact with the standard nonlinear realization formalism \cite{Callan:1969sn,Bando:1987br},
we can define a covariant differential operator
\bea
V_{\mu}= U^\dagger \partial_{\mu} U - i g U^\dagger  W_{\mu} U + i {g'} B_{\mu}\,,
\label{covdo}
\eea
therefore the mass term $ tr[D U^{\dagger} \cdot D U] $ can be formulated
as 
\bea
tr[D U^{\dagger} \cdot D U] &=& - tr[V \cdot V]\,.
\eea
The minus sign is from the fact that 
$(D_{\mu} U^{\dagger}) = - U^\dagger (D_{\mu} U) U^\dagger$.

Compared with the nonlinear sigma model,
we are fully aware of that there exist two types 
of power counting rules which can be introduced:
1) Weinberg's derivative power counting rule \cite{Weinberg:1978kz}. 
In order to make all these operators
as $O(p^2)$ order, we must assign
\bea
&&[v]_p=[h]_p=[W]_p=[B]_p=[\xi]_p=0\,,\,\,\nnb \\
&&[\partial]_p=[g]_p=[g']_p=1\,,\,\,[\mu^2]_p=[\lambda]_p=2\,.
\eea
This power counting rule is self-consistent in the
assignment of $[v]_p=0$, which is determined as $v=\sqrt{\mu^2/\lambda}$.
With this power counting rule, the one-loop 
divergence terms should be counted as $O(p^4)$,
which can be served as self-consistency checking 
for this rule. 
2) Georgi-Manohar mass dimension power 
counting rule \cite{Manohar:1983md} without Fermion is equivalent
to count the mass dimension
of fields and couplings, we have
\bea
&&[g]_m=[g']_m=[\lambda]_m=0\,,\,\,[v]_m=[h]_m=1\,,\nnb \\
&&[W]_m=[B]_m=[\xi]_m=[\partial]_m=1\,,\,\, [\mu^2]_m=2\,.
\eea
Due to the fact that the theory is renormalizable when formulated
with a linear form, so the one-loop divergence terms
are expected to just repeat those tree-level operators while the couplings and mass parameter
are expected to shift in order to absorb the UV divergences, which is exactly the meaning
of the renormalizability. 

These two power counting rules also serve as an important guidance 
for us to check our intermediate and final results.

\section{The nonlinearly realized SM in the background field method}

The Higgs sector beaks the local symmetry 
$SU_L(2) \times U_Y(1) \rightarrow U_{\textrm{EM}}(1)$ and
fixes the mixing in the vector sector
from weak interaction eigenstates to
the mass eigenstates.
The transformation from the weak eigenstates to
the mass eigenstates is determined by the 
following relations
\bea
W^\pm&=&{1\over \sqrt{2}} \left( W^{1} \mp i W^{2} \right)\,,\nnb\\
Z  &=&  \sin\theta_W B - \cos\theta_W W^{3} \,,\nnb\\
A  &=&  \cos\theta_W B + \sin\theta_W W^{3} \,,
\eea
where $\theta_W$ is Weinberg angle.

Correspondingly, the rotation in the vector fields induces
the corresponding rotation in the gauge group parameters
\bea
\al^\pm&=&{1\over \sqrt{2}} \left( \al_1 \mp i \al_2 \right)\,,\nnb\\
\al_Z  &=&  \sin\theta_W \beta_Y - \cos\theta_W \al_3 \,,\nnb\\
\al_A  &=&  \cos\theta_W \beta_Y + \sin\theta_W \al_3 \,.
\label{gpr}
\eea
We observe that these orthogonal rotations do 
not change neither the Lagrangian nor the functional
measure nontrivially. Here, the rotation in the gauge group parameters
is matched with the rotation of the vector
fields, which is necessary to guarantee the 
well-defined ghost sector. We will address the issue of 
matching in the next section.

In the spirit of the background field gauge 
quantization \cite{DeWitt:1967ub,DeWitt:1967uc,Gasser:1983yg}, we can decompose the Goldstone 
field into the classic part ${\overline U}$
and quantum part $\xi$ 
as
\bea
U\rightarrow {\overline U} {\widehat U}\,,\,\, {\widehat U} = \exp\{{i 2 \xi \over ( v+ \ch) }\}\,.
\label{upara}
\eea
Here we parameterize the Goldstone phase
in $\xi/(v + \ch)$. To parameterize the quantum Goldstone 
field in the above form is to simplify
the presentation of the standard form of quadratic terms of Goldstone bosons. 
With this parameterization, the D'Alambert operator of Goldstone bosons enjoys
the advantage that the quartic divergences are well-organized \cite{Dutta:2004te}.  

Compared with the standard parameterization of the Goldstone's degree of freedom in the
hadronic chiral Lagrangian \cite{Gasser:1983yg}, our parameterization modifies the definition of the
Goldstone 
\bea
\xi &=& \frac{v + \ch}{v} {\tilde \xi}\,,
\eea
where the ${\tilde \xi}$ is  the standard Goldstone field, or $\pi$ fields in the
hadronic chiral Lagrangian. This parameterization also induces a new quartic term
in the Lagrangian and modifies it as $\de^4(0) ln (1+ (\qh/(v + h)) )$. Such a modification in the
parameterization of the Goldstone fields 
is justified by the equivalence theorem \cite{Haag:1958vt}. However, 
such a parameterization of Goldstone fields becomes invalid when
the system has singular solutions with $v+ \ch=0$ (vortex solutions) \cite{Nielsen:1973cs}.
However, we only care about the perturbation and UV divergences around the regular solution
$v+\ch \neq 0$, therefore such a parameterization can be acceptable.

The vector boson field are split as
\bea
V_\mu \rightarrow {\overline V_\mu} + {\widehat V_\mu}\cma
\eea
where ${\overline V_\mu}=({\overline W_\mu}, {\overline B_\mu})$ represents the classic background
vector fields and ${\widehat V_\mu}=({\widehat W_\mu}, {\widehat B_\mu})$ represents the quantum
vector fields.

By using the Stueckelberg transformation 
\cite{Stuckelberg:1933,Grosse-Knetter:1992nn} for the
background vector fields,
\bea
&&{\overline W}^{s} \rightarrow {\overline U^{\dagger}} {\overline W} {\overline U} + i {\overline U}^{\dgr} \partial{\overline U} \cma
{\overline B^s} \rightarrow {\overline B}\cma \nnb \\
&& {\widehat W}^{s} \rightarrow {\overline U^{\dagger}} {\widehat W} {\overline U} \cma
{\widehat B^s} \rightarrow {\widehat B}\,,
\label{stueck}
\eea
the background Goldstone fields can be completely 
absorbed by redefining the background vector fields,
and will not appear in the one-loop effective Lagrangian.
By formulating the term $tr[D U^{\dagger} \cdot D U]$ into 
the standard nonlinear chiral realization
form $- tr[V \cdot V]$ \cite{Callan:1969sn,Bando:1987br}, we observe that in the
Stueckelberg transformation, only $SU(2)$ weak bosons are modified while
the $U(1)$ boson is unaffected.

The Stueckelberg fields is invariant under the
gauge transformation of the background gauge fields,
such a property guarantees that the following computation
is gauge invariant under the classic gauge transformation
from the beginning if we can express
all effective vertices into the Stueckelberg fields.
Such a fact is more apparently by examining the partition functional $Z$, which
is invariant under the classic gauge transformation.
After the loop calculation, by using the reversed
Stueckelberg transformation, the Lagrangian can be
restored back to the form represented by its low energy degree of freedom \cite{Grosse-Knetter:1992nn}:
the transverse vector bosons and the longitudinal Goldstone bosons.

In effect, the combination of the background Goldstone boson field and background
vector boson field into the Stueckelberg fields is equivalent to take 
the unitary gauge for the classic fields. In this case, the mass eigenstates
of $A$ and $Z$ are understood as the mixing between the third component
of the dressed $SU(2)$ vector fields and the undressed $U(1)$; the mass eigenstates of $W^\pm$
as the mixing between the first and second components of the dressed $SU(2)$ vector
fields. Since only the quantum fields need to be quantized, we can take different gauges for
the classic fields and the quantum fields \cite{DeWitt:1967ub,DeWitt:1967uc}, which is one of the advantages of the
background field method.

Similarly to the vector bosons, the Higgs boson is split as
\bea
h = {\overline h} + {\widehat h}\,,
\eea
where ${\overline h}$ and ${\widehat h}$ are classic and quantum Higgs
boson, respectively.

\section{the EOM, quantization and the gauge fixing terms}

The equation of motion of the background classic 
vector fields (Stueckelberg field, below we 
omit the $^s$ in both the classic and quantum 
fields) is determined by the Euler-Lagrange variation
method, which can be simply formulated as
\bea
D_{\mu} {\cal \widetilde W}^{\nu\mu} = - \sigma_{0,VV} V^{\nu}\,,
\eea
with ${\cal \widetilde W}^{\mu\nu,T}=\{ A^{\mu\nu} + i e F_Z^{\mu\nu},
Z^{\mn} - i {g^2 \over G} F_Z^{\mn},
 W^{+,\mn} -i g F^{+,\mu\nu} , W^{-,\mn} - i g F^{-, \mu \nu} \}$. Here
and below, we omit the overline on the classic field for the simplicity of
representation.
The $V^T$ means the $(A, Z, W^+, W^-)$.
The relevant definition are given as
\bea
&A_{\mu \nu}  =& \partial_{\mu} A_{\nu} - \partial_{\nu} A_{\mu}\,,\nnb \\
&Z_{\mu \nu}  =& \partial_{\mu} Z_{\nu} - \partial_{\nu} Z_{\mu}\,,\nnb  \\
&W^{\pm}_{\mu \nu}  =& d_{\mu} W_{\nu}^\pm - d_{\nu} W^\pm_{\mu}\,,\nnb \\
&F_Z^{\mu\nu}=& W^{\mu}_+ W^{\nu}_- - W^{\mu}_- W^{\nu}_+ \,,\nnb \\
&F_\pm^{\mu\nu} =& \pm (Z^{\mu} W^{\nu}_\pm - W^{\mu}_\pm Z^{\nu})\,, \nnb \\
&d_\mu W_{\nu}^\pm =& \partial_\mu W^\pm_\nu - i (\pm 1 ) e A_\mu W_\nu^\pm\,.
\eea
The covariant operator $D$ will be given when we define the D'Alambert operator
of vector boson fields.

The EOM of vector bosons
induces the following relations
\bea
\partial \ln (v + \ch) \cdot Z &=& - { 1\over 2} \partial \cdot Z\,,\label{eom1}\\
\partial \ln (v + \ch) \cdot W^{+} &=& - {1 \over 2} d \cdot W^{+} + i {1 \over 2} {g'^2\over G} Z \cdot W^{+}\,,\\
\partial \ln (v + \ch) \cdot W^{-} &=& - {1 \over 2} d \cdot W^{-} - i {1 \over 2} {g'^2\over G} Z \cdot W^{-}\,,\label{eom2}
\eea
which can also be obtained by extracting the linear terms of Goldstone fields.
According to the our calculation, it seems that it is not necessary to
impose the condition $\partial A=0$.

The equation of motion of the background Higgs field is given as
\bea
\partial^2 \ch &=& ( v+\ch) [{G^2 \over 4} Z \cdot Z + {g^2 \over 2} W^{+} \cdot W^{-}
\wsep - \mu^2 + {\lambda \over 4} ( v+ \ch)^2 ]\,.
\label{eomhiggs}
\eea
These equations of motion satisfy both power counting rules. In the Weinberg's
derivative power counting rule, both sides of the EOM of vector and Higgs fields are counted
as  $2$. In the Georgi-Manohar mass dimension power counting rule, 
both sides of the EOM of vector and Higgs fields are counted as $3$ (The EOM of
photon is different, only left hand side is counted as $3$, while the right hand side
is counted as $0$).

The quantization of the dynamics system is selected to be performed  
around the classic gauge transformation given in Eq. (\ref{stueck}). 
The quantum degree of freedoms include quantum vector bosons, quantum
Goldstone bosons, and quantum Higgs bosons, and the partition functional without
gauge fixing terms can be formulated as
\bea
Z[V, h]={\cal N} \int {\cal D} {\cal \al} {\cal D} {\widehat V} {\cal D} {\widehat h}  {\cal D} {\xi} \cdots \,,
\eea
where the vector and Goldstone bosons are formulated in mass eigenstates.
The vector quantum fields, ${\widehat V}$, represent ${\widehat A}$, ${\widehat Z}$,
${\widehat W^+}$, and ${\widehat W^-}$.
In the Goldstone bosons, unlike in the gauged nonlinear $SU(2)$ sigma model, there
is no $SU_c(2)$ global symmetry any more. And the Goldstone bosons must be
parameterized as $\xi_Z$, $\xi_+$, and $\xi_-$.

In order to separate the infinite volume of gauge group parameter, we use the
covariant background gauge fixing \cite{DeWitt:1967ub,DeWitt:1967uc}. 
Follow the standard quantization procedure, the infinite gauge parameter
volume is separated by imposing gauge fixing to the quantum fluctuations, as
given as
\bea
Z[V, h]=\int {\cal D} {\cal \al} Z_G[V, h]\,,
\eea
where $Z_G$ is the fixed partition functional, which can be expressed
as 
\bea
Z_G[V, h]={\cal N'} \int {\cal D} {\widehat V} {\cal D} B {\cal D} {\widehat h}  {\cal D} {\xi} {\cal D} {v} {\cal D} {u} \cdots
\eea 
Quantization procedure introduces
two types of real ghost fields, $v$ and $u$, into the quantum 
degree of freedoms. Here we follow the
convention of references \cite{Alkofer:2000wg,Kugo:1979gm,Nakanishi:1990qm} to label these two
types of ghost differently.

The Lagrangian of $Z_G$ is given as
\bea
{\cal L}&=&{\cal L}(V, \ch; {\widehat V},\xi, {\widehat h}) + {\cal L}_{NL}\,,
\eea
where the first term contains the classic vector bosons (in U-gauge) and the classic
Higgs bosons, the quantum fields. And the second term contains gauge fixing terms.
In the Nakanishi-Lautrup auxiliary field form,
the gauge fixing and ghost terms \cite{Becchi:1975nq} can be formulated as 
\bea
\label{lnl}
{\cal L}_{NL} & =& i F^a B^a - \frac{1}{2} B^a \Gamma^{ab} B^b  
\wsep -  i v^a M^{ab} \frac{\de F^b}{\de \al^{\bar a}} {\overline M^{{\bar a} {\bar b}}} u^{\bar b}\,.
\eea
Here we have deliberately used the index ${\bar b}$ to label
$u$ type ghost and the gauge group parameter index 
to distinguish the ghost and anti-ghost. Due to the
structure of the semi-simple group, the matrices $\Gamma^{ab}$,
$M^{ab}$, and $M^{{\bar a} {\bar b}}$, are introduced to accommodate the
rotation of $F^a$, $v^a$ and $u^{\bar a}$. 
The matrices $M^{ab}$ and ${\overline M^{{\bar a} {\bar b}}}$ are related with
the ambiguity in exponentializing the Faddeev-Popov determinant to the
Lagrangian. Such an
ambiguity could occur if the $\det M^{ab}$ and $\det {\overline M^{{\bar a} {\bar b}}}$ do
not vanish, the following relation can be justified 
\bea
\det(\frac{\de F^a}{ \de \al^{\bar b}})&=& \frac{ det( M^{ab} \frac{\de F^b}{\de \al^{\bar a}} {\overline M^{{\bar a}{\bar b}}})}
{\det(M^{ab}) \det({\overline M^{{\bar a}{\bar b}}})}\,.
\eea
If the $F^a$ (here the index $a$ is related with quantum field index)
rotatate to $a'$, in order to have well-defined canonical ghost fields,
the $\al^{\bar a}$ must rotate accordingly. For example,
for the gauge fixing given in Eq. (\ref{gfx1}, where the
field index have changed from $a=(Y,3,1,2)$ to 
$a'=(A,Z,W^+,W^-)$, if we do not
defined $\al^A$ as given in Eq. (\ref{gpr}) while keep using
the $\al_Y$ and $\al_3$, we will find that the ghost kinetic term
have the following form
\bea
& -  v^A \partial^2 (\cos\theta_W u^{Y} + \sin\theta_W u^{3})  \nnb \\
& -  v^Z \partial^2 (\sin\theta_W u^{Y} - \cos\theta_W u^{3}) 
+ \cdots\,.
\eea
In the Hamiltonian form with the canonical ghost fields
and its conjugate fields, and the canonical commutation
relation and the propagators can not be well-defined, 
therefore the quantization procedure has problems. 
So we emphasize here, that the matching
between the rotations of gauge fields and gauge group parameters
are important for the ghost term.

To represent ghost
fields and anti-ghost differently is for the correct
hermiticity assignment as pointed out by \cite{Kugo:1979gm,Nakanishi:1990qm}, 
and is well-known in the conformal field theory \cite{conformal} where
$b$ and $c$ type real ghosts are introduced.
Here we would like to emphasize that the rotations of
ghost and anti-ghost are independent. The rotation of
$v$ type ghosts can be related with the change of
$F^a$ fields, and the rotation of $u$ type ghosts
are related with the rotation of the gauge parameter $\al$.
After diagonalizing of Nakanishi-Lautrup auxiliary fields $B^a$,
the eigenvalues of the matrix $\Gamma$ are 
required to be  real and positive. The Landau gauge and the unitary gauge,
which correspond to the case that some of the eigenvalues of the matrix
$\Gamma$ is zero and infinite, respectively. These two extreme cases
should be treated carefully in the functional integral, as shown in \cite{Baulieu:1981sb} and in \cite{Grosse-Knetter:1992nn},
respectively.

The quantization of Yang-Mills fields 
can be performed in a more strict and
sophisticated way by incorporating
BRST global symmetry in the formalism, as shown
in \cite{Gomis:1994he}. However, for our purpose,
here we provide a simple and direct
understanding on the quantization, the gauge fixing, and
gauge parameter independence for the S-matrix.
The quantization and gauge fixing 
can be understood as a procedure to couple 
the vector dynamic system to an auxiliary 
system. Even after the covariant Lorentz gauge fixing, 
in the $SU(N)$ theories for instance, the auxiliary system can still enjoy
an isospin-like global symmetry and scaling
symmetry as analyzed in \cite{'tHooft:1976fv}. The global symmetry and scaling
symmetry can be equally represented in the arbitrariness of choosing
$\Gamma$ and two $M$ matrices in the ${\cal L}_{NL}$.
The S-matrix should be independent of parameters of these matrices.

The real ghost fields enjoy a scaling transformation invariance
which is related with the ghost number conservation law.
\bea
 v &\rightarrow e^{\lambda} v \nnb \\
 u &\rightarrow e^{-\lambda} u \,,
\eea
$v$ is identified as anti-ghost field, while $u$ is identified as ghost
field. This ghost number conservation scaling transformation is consistent with
the hermiticity assignment \cite{Kugo:1979gm,Nakanishi:1990qm}.
This scaling group is the subgroup of 
general matrix $M$ and ${\overline M}$,
both of which are defined with each component is a complex number.

The gauge fixing term for the quantum fields are chosen in the mass eigenstates, and are given
as below 
\bea
\label{gfx1}
{\cal L}_{GF,A}&=&-{1 \over 2} (\partial\cdot\hA  \wsep - i e (\hWm \cdot \bWp - \hWp \cdot \bWm))^2\cma\\
{\cal L}_{GF,Z}&=&-{1 \over 2} (\partial\cdot\hZ - {1\over 2} G (v + \ch)  \xi_Z
\wsep + i {g^2 \over G} (\hWm \cdot \bWp - \hWp \cdot \bWm))^2\cma\\
{\cal L}_{GF,W}&=&- (d\cdot\hWp + {1\over 2} g ( v + \ch) \xi_W^+
      + i {g^2 \over G} \bZ \cdot \hWp \wsep - i {g^2 \over G} \bWp \cdot \hZ
	+ i e \bWp \cdot \hA)\nnb\\&&(d\cdot\hWm + {1\over 2} g ( v + \ch) \xi_W^-
      - i {g^2 \over G} \bZ \cdot \hWm \wsep + i {g^2 \over G} \bWm \cdot \hZ
	- i e \bWm \cdot \hA)\,.
\label{gfx2}
\eea
To determine the gauge parameters, we use the following guidances: 
1) The D'Alambert operator of quantum vector field must have the
form as given in Eq. (\ref{op1}). When gauge potential is neglected,
this correspond to the Feynman-'t Hooft gauge in the momentum space.
The terms with background fields in the gauge fixing term have the function
to eliminate extra Lorentz structures in the vector sector, and guarantee
a well-defined exact vector propagator in coordinate space.
2) Although we can introduce nonlinear terms of $\xi$, up to one-loop level,
we only consider the linear term in the gauge fixing. This linear term induces
mass term of the corresponding ghost fields. The nonlinear terms might be useful
when we consider beyond one-loop calculation, and these nonlinear terms also
induce nonlinear interactions between Goldstone and ghost. Since we only consider
one-loop renormalization, we confine to the linear term.
3) There is no mixing between massive vector fields and their corresponding Goldstone bosons,
and there is no mixing between massless photon fields and chargeless Goldstone $\xi_Z$;
which is a typical feature of Feynman-'t Hooft gauge.
4) The classic $U_{\textrm{EM}}(1)$ symmetry must be fulfilled explicitly.

These gauge fixing terms will fulfill two important functions when we consider to
extract divergences: 
1) to guarantee that the exact vector propagator have well-controlled divergence structure.
2) to eliminate the kinetic mixing term between the
vector and Goldstone fields and to make the expansion of the vector and Goldstone mixing term 
$Tr\ln( 1 - \rx_{\xi} \Box_{VV}^{-1} \lx_{\xi} \Box_{\xi\xi}^{-1})$
to stop in producing UV divergences to a specific term. 
We avoid in using the Landau gauge as being used in \cite{Longhitano:1980iz,Longhitano:1980tm,Herrero:1993nc,Herrero:1994iu,Dittmaier:1995ee}, 
which has the advantage to identify the massless poles to the Goldstone particles
in the asymptotic region and to eliminate the ghost-Goldstone interactions. The reason is
that it makes the expansion of 
vector-Goldstone mixing term complicated. 

By using the arbitrariness in choosing $\Gamma$ and two $M$ matrices, we can
rotate these gauge fixing terms back to weak interaction form (the background
vector fields and gauge group parameters also rotate accordingly), 
which have the following form
\bea
\label{gfx1w}
F_Y &=& \partial \cdot {\hat B} - \frac{g'}{2} (v + h) \xi^Z \,,\nnb\\
F^i_W &=& (\de^{ij} \partial + g f^{ijk} W^{j}) 
\cdot {\hat W}^{k}  + \frac{g}{2} (v + h )  \xi^i\,.
\label{gfx2w}
\eea
Such a fact demonstrates we can put gauge fixing terms either in mass eigenstates
or in weak interaction eigenstates for vector bosons, and they are equivalent with each other
and do not affect the S-matrix, as indicated by $\Gamma$ and two $M$ matrices in 
the ${\cal L}_{NL}$ given in Eq. (\ref{lnl}).
Gauge fixing terms in the weak interaction eigenstate
basis have the advantage to explicitly show that our gauge fixing terms are
covariant background field gauge fixings.

\section{The quadratic forms of the one-loop Lagrangian}

Since we only care about the one-loop renormalization, we can only concentrate
on quadratic terms (bilinear terms) of quantum fields and neglect
those trilinear terms, quartic terms, and higher order terms in the Lagrangian.
Those terms neglected are necessary when we consider higher order renormalization.

In order to make the one-loop renormalization easier to handle,
we cast the quadratic quantum fluctuation terms of the one-loop Lagrangian
into a standard form, which reads
\bea
{\cal L}_{quad}&=&- \bbbkf{
{1\over 2 } {\widehat V_{\mu}^{\dagger a}} \Box^{\mu\nu,ab}_{V\,V} {\widehat V_{\nu}^b}
+ {1\over 2} \xi^{\dagger i}  \Box_{\xi\,\xi}^{ij} \xi^j
\wsep + v^a  \Box_{vu}^{ab} u^b
+ {1\over 2} {\hat h} \Box_{hh} {\hat h}\wsep
+ {1\over 2} {\widehat V_{\mu}^{\dagger,a}} {\stackrel{\leftharpoonup}{X}}_{\xi}^{\mu,aj} \xi^j
+ {1\over 2} \xi^{\dagger,i} {\stackrel{\rightharpoonup}{X}}_{\xi}^{\nu,ib} {\widehat V_{\nu}^b}
\wsep+ {1\over 2} {\widehat V_{\mu}^{\dagger,a}} {\stackrel{\leftharpoonup}{X}}_{h}^{\mu,a} {\widehat h}
+ {1\over 2} {\widehat h} {\stackrel{\rightharpoonup}{X}}_{h}^{\mu,a} {\widehat V_{\mu}^{a}}
\wsep + {1\over 2} \xi^{\dagger,i} X_{\xi h}^i {\widehat h}
+ {1\over 2} {\widehat h} X_{h \xi}^i \xi^{i}
}\,,
\label{lag}
\eea
\bea
\Box^{\mu\nu,ab}_{V\, V} &=& - D^{2,ab} g^{\mu\nu} + \sigma_{0,VV}^{ab} g^{\mu\nu} + \sigma_{2,VV}^{\mu\nu,ab}\,\,, \label{op1}\\
\Box_{\xi\,\xi}^{ij}&=& - d^{2,ij} +  \sigma_{0,\xi\xi}^{ij}  + \sigma_{2,\xi\xi}^{ij}\,,\\
\Box_{hh} &=& - \partial^2 + m^2_h + \sigma_{h h}\,,\\
\Box_{vu}^{ab}&=& - D^{2,ab} + \sigma_{0,VV}^{,ab}\,, \label{op2}\\
X_{h \xi}^i &=& X_{h \xi}^{\alpha, i} d_{\alpha}        + X_{h \xi,0}^{i}\,,\\
X_{\xi h}^i &=& X_{\xi h}^{\alpha, i} \partial_{\alpha} + X_{\xi h,0}^{i}\,,
\label{stdfesm}
\eea
where ${\widehat V^{\dagger}}=({\widehat A},{\widehat Z},{\widehat W^{-}},{\widehat W^{+}})$ 
and ${\widehat V^T} = ({\widehat A},{\widehat Z}, {\widehat W^{+}}, \widehat{W^{-}})$,
$\xi^{\dagger}=(\xi_Z, \xi^{-}, \xi^{+})$ and $\xi^T=(\xi_Z, \xi^{+}, \xi^{-})$,
and $v = (v_A, v_Z, v_-, v_+, )$ and $u^T = (u_A, u_Z, u_+, u_-)$.
The Hermitian transformation 
in our calculation is defined as
\bea
(V)^\dagger = V^{\dagger}\,,\,\,(\xi)^\dagger = \xi^\dagger\,,\,\,
(v)^{\dagger} = {v^{\dgr}}\,,\,\,(u)^{\dagger}={u^\dgr}\,.
\eea
The Hermitian transformation is equal to the charge conjugate transformation
in the vector, Goldstone, and ghost sector.
The Lagrangian given in Eq. (\ref{lag}) is invariant under 
this Hermitian transformation (charge conjugate). 
Here we would like to emphasize that our Hermitian transformation
has only charge conjugate and does not require the transpose transformation.

The covariant differential operators $D=\partial+\Gamma_V$
and $d=\partial+\Gamma_{\xi}$, and the gauge connection of
vector bosons $\Gamma_V$ is defined as
\begin{displaymath}
\left (\begin{array}{cccc}
0&0& i e W^{-}_{\mu}&-i e W^{+}_{\mu}\\
0&0&-i {g^2 \over G} W^{-}_{\mu} & i {g^2 \over G} W^{+}_{\mu}\\
 i e W^{+}_{\mu} & - i {g^2 \over G} W^{+}_{\mu}& - i e A_{\mu} + i {g^2 \over G} Z_{\mu} &0\\
-i e W^{-}_{\mu} &   i {g^2 \over G} W^{-}_{\mu}&0&i e A_{\mu} - i {g^2 \over G} Z_{\mu}
\end{array}\right)\,.
\end{displaymath}
The gauge connection of Goldstone bosons $\Gamma_{\xi}$ is defined as
\begin{displaymath}
\left (\begin{array}{ccc}
0&i {g\over 2} W^{-}_{\mu}&-i {g\over 2} W^{+}_{\mu}\\
i {g\over 2} W^{+}_{\mu}& - i e A_{\mu}  &0\\
- i {g\over 2} W^{-}_{\mu}&0&i e A_{\mu}
\end{array}\right)\,.
\end{displaymath}

While for the ghost term the covariant differential operator is the same one as that of
the vector boson fields, similar to the case without spontaneous symmetry breaking.

All gauge potentials satisfy the anti-Hermitian property like $\partial$, therefore
the differential operators $i D_{\mu}$ is Hermitian. 
All exact propagators in Eqs. (\ref{op1}---\ref{op2})
are Hermitian, so their spectra are expected to be positive and definite and the heat kernel
regularization (and the Schwinger proper time) method can be reliably 
used to extract divergences. Here due to the fact that there is no Fermion included,
we do not need to use any extra prescription to make the operators Hermitian, as people do 
for the most general Dirac operator which violates the hermiticity assignment \cite{Gasser:1983yg}.

Here we would like to comment on the hermiticity assignment given in \cite{Kugo:1979gm,Nakanishi:1990qm}.
The ghost fields are scalar fields but satisfy anticommutation
relation. In order to construct physical spectrum of the non-Abelian theory and have
a well-defined S-matrix in the operator formalism, 
the correct hermiticity assignment should be defined,
such a convention is widely taken in conformal and string
theories. Considering the
remnant $U_{\textrm{EM}}(1)$ symmetry, here we have modified this hermiticity
assignment simply as charge conjugate. Historically, 
the ghost fields are introduced as complex fields. 
For the perturbation calculation, for all existed 
examples as far as we know, such an assignment makes no difference. 
However, we find
when $O(p^4)$ operators are included, such a hermiticity
assignment indeed makes difference \cite{preparation}. 
Anyhow, for the current two cases, such a hermiticity assignment
makes no difference.

The matrix $\sigma_{2,VV}$ contains Lorentz tensor structure and 
corresponds to the dipole magnetic term of vector bosons due to the
fact that vector bosons are spin-one particles.
It is given as
\begin{displaymath}
\sigma_{2,VV}^{\mu\nu,ab} =\left (\begin{array}{cccc}
0&0&\sigma_{2,AW^+}^{\mu\nu} &\sigma_{2,AW^-}^{\mu\nu} \\
0&0&\sigma_{2,ZW^+}^{\mu\nu} &\sigma_{2,ZW^-}^{\mu\nu} \\
\sigma_{2,W^-A}^{\mu\nu} &\sigma_{2,W^-Z}^{\mu\nu}  &\sigma_{2,W^-W^+}^{\mu\nu}  &0 \\
\sigma_{2,W^+A}^{\mu\nu} &\sigma_{2,W^+Z}^{\mu\nu}  &0  &\sigma_{2,W^+W^-}^{\mu\nu}
\end{array}\right)\,,
\end{displaymath}
and each of its components reads as
\bea
\sigma_{2,AW^+}^{\mu\nu}&=&-\sigma_{2,W^+A}^{\mu\nu} =   2 i e {\widetilde W^{-,\mu\nu}}\,,\nnb\\
\sigma_{2,AW^-}^{\mu\nu}&=&-\sigma_{2,W^-A}^{\mu\nu} = - 2 i e {\widetilde W^{+,\mu\nu}}\,,\nnb\\
\sigma_{2,ZW^+}^{\mu\nu}&=&-\sigma_{2,W^+Z}^{\mu\nu} = - 2 i {g^2 \over G} {\widetilde W^{-,\mu\nu}}\,,\nnb\\
\sigma_{2,ZW^-}^{\mu\nu}&=&-\sigma_{2,W^-Z}^{\mu\nu} =   2 i {g^2 \over G} {\widetilde W^{+,\mu\nu}}\,,\nnb\\
\sigma_{2,W^- W^+} &=& - \sigma_{2,W^+ W^-} = 2 i g W^{3,\mu\nu}\,.
\eea
Here we would like to mark that $\sigma_{VV}$ is Hermitian. Under the Hermitian transformation,
the Lorentz indices should be exchanged as $\mu\rightarrow\nu(\nu\rightarrow\mu)$, and this
makes the invariance of $\sigma_{VV}$ explicitly. The $\sigma_{VV}$  does not depend on the 
background Higgs field $\ch$.

Goldstone bosons are spin-zero particles, but the matrix $\sigma_{2,\xi\xi}$ does not vanish
and has scalar structure in the Lorentz index. It is given as
\begin{displaymath}
\sigma_{2,\xi\xi}^{ij} =\left (\begin{array}{ccc}
\sigma_{2,\xi_Z\xi_Z}& \sigma_{2,\xi_Z\xi^+} & \sigma_{2,\xi_Z\xi^-}\\
\sigma_{2,\xi^-\xi_Z}& \sigma_{2,\xi^-\xi^+} &\sigma_{2,\xi^-\xi^-}\\
\sigma_{2,\xi^+\xi_Z}& \sigma_{2,\xi^+\xi^+} &\sigma_{2,\xi^+\xi^-}
\end{array}\right)\,,
\end{displaymath}
and each of its components reads as
\bea
\sigma_{2,\xi_Z\xi_Z}&=& \frac{g^2}{2} W^{+} \cdot W^{-} - \frac{ \partial^2 h } {v+ h}\,,\nnb\\
\sigma_{2,\xi^+\xi^+}&=&- {g^2 \over 4} W^{-} \cdot W^{-}\,,\nnb\\
\sigma_{2,\xi^-\xi^-}&=&- {g^2 \over 4} W^{+} \cdot W^{+}\,,\nnb\\
\sigma_{2,\xi_Z\xi^+}&=&\sigma_{2,\xi^+\xi_Z}={g G \over 4} W^{-} \cdot Z\,,\nnb\\
\sigma_{2,\xi_Z\xi^-}&=&\sigma_{2,\xi^-\xi_Z}={g G \over 4} W^{+} \cdot Z\,,\nnb\\
\sigma_{2,\xi^+\xi^-}&=&\sigma_{2,\xi^-\xi^+}= \frac{g^2}{4} W^{+} \cdot W^{-} + \frac{G^2}{4} Z \cdot Z\wsep - \frac{ \partial^2 h} {v+ h}\,.   
\eea
The $\sigma_{2,\xi\xi}$, under the Hermitian transformation,
is explicitly invariant. The term $\frac{ \partial^2 h } {v+ h}$
in the $\sigma_{2,\xi_Z\xi_Z}$ and $\sigma_{2,\xi^+\xi^-}$ is not trivial, and
it represents a nonperturbative expansion on the $\ch$ field.  By using the
momentum expansion and simply extracting Feynman rule, we can not obtain such
a compact form.

In order to explicitly trace the function of the
EOM of the Higgs field to eliminating extra divergence structures, we
have restricted from substituting $\partial^2 h$ term in $\sigma_{2, \xi \xi}$
by the equation Eq. (\ref{eomhiggs}).

The ghost fields are spin-zero particles, 
their propagators have no nontrivial $\sigma$ structure, like ordinary scalar particles. 
However, we observe that the ghost fields are different from the Goldstone fields in this
aspect.

The $\sigma_{hh}$ is determined as
\bea
\sigma_{hh} &=&-{1\over 4} (G^2 Z\cdot Z + 2 g^2 W^+ \cdot W^-) 
\wsep - {\lambda \over 4} v^2 - {3\over 4} \lambda ( v+ \ch)^2\,.
\eea
And the Hermitian transformation  invariance is apparently.  We observe that
the $\sigma_{hh}$ depends on the background vector fields.

The mass matrices have the form $\sigma_{0, VV}^{ab}=dia\{0,G^2 {(v+\ch)^2}/{4},g^2 {(v+\ch)^2}/{4},g^2 {(v+\ch)^2}/{4}\}$
and $\sigma_{0,\xi\xi}^{ij}=dia\{G^2 {(v+\ch)^2}/{4},g^2 {(v+\ch)^2}/{4},g^2 {(v+\ch)^2}/{4}\}$.
The background Higgs field $\ch$ contributes to the mass matrices.
Due to the gauge fixing term we have chosen, we observe that the vector boson, 
its corresponding Goldstone and
ghost fields, have the same mass term, similar to the Feynman-'t Hooft gauge.
The $m^2_h$ is the mass of Higgs boson, which is determined as $m^2_h=\lambda v^2/2$.

The mixing term between the vector and Goldstone bosons
is a $4\times 3$ matrix, and is determined as
\bwt
\begin{displaymath}
{\stackrel{\leftharpoonup}{X}}_{\xi}^{\mu,aj}  =\left (\begin{array}{ccc}
0&-i {g^2 g' \over G} (v + h) W^{-,\mu} & i {g^2 g' \over G} (v + h) W^{+,\mu}\\
G \partial^{\mu} h & i {g \over 2 G} (g^2 - g^{'2}) (v + h) W^{-,\mu}& - i {g \over 2 G} (g^2 - g^{'2}) (v + {h}) W^{+,\mu}\\
- i {g^2 \over 2} ( v + h) W^{+,\mu}&-g \partial^{\mu} h - {1\over 2} i g G (v+ h) Z^{\mu}&0\\
  i {g^2 \over 2} ( v + h) W^{-,\mu}&0&-g \partial^{\mu} h + {1\over 2} i g G (v+ h) Z^{\mu}
\end{array}\right)\,,
\end{displaymath}\ewt
while the matrix ${\stackrel{\rightharpoonup}{X}}_{\xi}^{\mu,aj}$ is a $3\times4$ matrix and
is just
the rearrangement of the ${\stackrel{\leftharpoonup}{X}}_{\xi}^{\mu,aj}$ due to our Hermitian
transformation (charge conjugate and transpose transformation), and
here we do not rewrite it explicitly.
The mixing term between vector and Higgs bosons
is determined as
\bea
{\stackrel{\leftharpoonup}{X}}_{h}^{\mu,a} &=&\{0,-{1\over 2} G^2 (v + h) Z^{\mu},\wsep
-{1\over 2} g^2 (v + h) W^{+,\mu},-{1\over 2} g^2 (v + h) W^{-,\mu}\}\,.
\eea
The mixing term ${\stackrel{\rightharpoonup}{X}}^{\mu,a}_h$
is the rearrangement of the ${\stackrel{\leftharpoonup}{X}}^{\mu,a}_h$ and is omitted here.
The mixing terms between Higgs and Goldstone bosons
are determined as
\bea
X_{h \xi}^{\alpha, i}&=& \{-G Z^{\alpha}, g W^{-,\alpha}, g W^{+,\alpha}\}\,,\\
X_{h \xi,0}^{i}      &=& \{-{G \over 2} \partial \cdot Z,
{g \over 2} d \cdot W^- - i {1\over 2} ({g^2 \over G} - G) W^- \cdot Z,
\wsep {g \over 2} d \cdot W^+ + i {1 \over 2} ({g^2 \over G} - G) W^+ \cdot Z \}.
\eea
The terms $X_{\xi h}^{\alpha, i}$ and $X_{\xi h,0}^{i}$ are also the rearrangement
of $X_{h \xi}$ and $X_{h \xi,0}$, and are omitted here. To extract
the mixing terms between Higgs and Goldstone bosons, we have used
the EOM of vector fields given in Eqs. (\ref{eom1}---\ref{eom2}).

It is an interesting fact that in both Weinberg's derivative power counting rule and Georgi-Manohar's power counting rule,
those propagators and mixing matrices given in Eq. (\ref{lag}) are counted as $2$; 

There is one remarkable feature about the ghost term: in the Feynman-'t Hooft
gauge we take, the $U_{\textrm{EM}}(1)$ ghost does not decouple from interaction with other
ghost fields. This is different from the common knowledge with pure Abelian gauge theories where
ghosts are expected to decouple. The underlying reason is
that the $U_{\textrm{EM}}(1)$ ghost field is a mixture of the $U_Y(1)$ and $SU_L(2)$ ghosts
and it couples to charged ghost fields while compatible with the principle 
of gauge invariance. Such a nondecouple behavior also help to cure the ambiguity
in the gauge fixing when a theory with a semi-simple group symmetry has
$U(1)$ groups \cite{Becchi:1975nq}.

\section{Evaluating the traces and logarithms}
By diagonalizing the quantum vector fields, Goldstone fields, and Higgs fields successively,                             
we can integrate out all quantum fluctuations
by using the Gaussian integral \cite{Dittmaier:1995ee}. And the ${\cal L}_{1-loop}$
can be concisely expressed as the traces and logarithms
\bea
Tr\Box_{vu} - {1\over 2} \mbra{ Tr\ln\Box_{VV}
+Tr\ln\Box_{\xi\xi}' + Tr\ln\Box_{hh}''}\,\,,
\label{logtrsm}
\eea
where
\bea
\Box_{\xi\xi}^{'ij}& =& \Box_{\xi\xi}^{ij} - \rx_{\xi} \Box_{VV}^{-1} \lx_{\xi}\,,
\eea
\bea
\Box_{hh}'&=&\Box_{hh} - \rx_h \Box_{VV}^{-1} \lx_h \,,
\eea
\bea
\Box_{hh}''&=&\Box_{hh}' - X'_{h\xi} \Box_{\xi\xi}^{'-1} X_{\xi h}\,,
\eea
\bea
X'_{h\xi} &=& X_{h\xi} - \rx_h \Box_{VV}^{-1} \lx_{\xi}\,,\\
X'_{\xi h} &=& X_{\xi h} - \rx_{\xi} \Box_{VV}^{-1} \lx_h\,.
\eea
Terms $Tr\ln\Box_{\xi\xi}'$ and $Tr\ln\Box_{hh}''$ are dependent on the
sequence as how to integrate quantum fields, and are determined by the following
relations
\bea
Tr\ln\Box_{\xi\xi}' &=& Tr\ln\Box_{\xi\xi}
\wsep + Tr\ln( 1 - \rx_{\xi} \Box_{VV}^{-1} \lx_{\xi} \Box_{\xi\xi}^{-1})\,,\\
Tr\ln\Box_{hh}''    &=& Tr\ln\Box_{hh}'
\wsep + Tr\ln( 1 - X'_{h\xi} \Box_{\xi\xi}^{'-1} X_{\xi h}  \Box_{hh}^{'-1})\,.
\eea
In order to extract all divergences at one-loop level, we need to expand the
Eq. (\ref{logtrsm}) to exact propagator terms and interaction terms.
Thanks to the gauge fixing terms given in Eqs. (\ref{gfx1}---\ref{gfx2}), 
we find that only the following terms can contribute the divergences \bwt
\bea
\int_x {\cal L}_{1-loop}&=& Tr\Box_{\bar{c}c}  - {1\over 2} \mbra{ Tr\ln\Box_{VV}
+Tr\ln\Box_{\xi\xi}  + Tr\ln\Box_{hh}
\nwsep - Tr(\rx_{\xi} \Box_{VV}^{-1} \lx_{\xi} \Box_{\xi\xi}^{-1})
 \nwsep- Tr(\rx_h \Box_{VV}^{-1} \lx_h \Box_{hh}^{-1})
\wsep  - Tr(X_{h\xi} \Box_{\xi\xi}^{-1} X_{\xi h} \Box_{hh}^{-1})
\nwsep - {1\over 2} Tr(X_{h\xi} \Box_{\xi\xi}^{-1} X_{\xi h} \Box_{hh}^{-1}X_{h\xi} \Box_{\xi\xi}^{-1} X_{\xi h} \Box_{hh}^{-1})
 + \cdots}
\,.
\label{divt}
\eea
\ewt
Here the $Tr$  means to sum over space-time points, Lorentz indices, 
and group indices, respectively.
Thanks to the property of the $Tr$, the above
equation clearly shows that divergences are independent of the sequence of 
integrating-out quantum fields. Even when finite terms are considered, this is also true.
This is a pleasant result which confirms that the sequence to integrate out
quantum fields has no physical meaning.

Since we only care about divergence structure, and we omit the finite terms by $\cdots$.
We would like to remark that each term in Eq. (\ref{divt}) corresponds 
to a series of Feynman diagrams. In principle, we can make a correspondence between Feynman diagrams
and terms in Eq. (\ref{divt}), though it is cumbersome.

In both power counting rule, these terms can be counted as $0$. 
So we must extract divergence
structure from these terms, which should be counted 
as $p^4$ terms in the Weinberg's power counting
rule and dimension $4$ in the Georgi-Manohar power counting rule.

\section{Divergences and the renormalizability at one-loop level}
To extract divergences, we need to introduce the third counting rule to
count the superficial divergence degree of freedom, so that we can know which terms
in the traces and logarithms contain divergences. 
When we calculate with the Schwinger proper time method in
coordinate space,  we introduce the following superficial divergence power counting rule
\bea
[\partial_\mu]_d=-1,\,\,[x_{\mu}]_d=1, \,\,[\tau]_d=2, 
\eea
where the $\partial_\mu$ means the differential partial operator which
might appear in the vertices,
the $x_{\mu}$ means the vector in the coordinate space, the $\tau$ means the
Schwinger proper time.  The total divergence degree of a term depends on
the number of integrals over the coordinate space, $n_x$, the number of proper time
integrals, $n_{\tau}$, the power index of the differential partial operators, $v_d$,
the power index of coordinate vectors, $v_x$, and the power index of
proper times, $v_{\tau}$. 
For example, the number of integrals over the coordinate space $\int_x \int_y$ is
counted as $2$, the number of integrals over the proper time
integrals $\int_\tau$ is counted as $1$, the power index of $\partial_x \partial_y$
is counted as $2$, the power index of $x_1^{\mu} x_2^{\nu}$ is 
counted as $2$, while the power index of $\tau_1 \tau_2^2$ is counted as $3$.
At one-loop, after expressing each term given in Eq. (\ref{divt}
with exact propagators in the coordinate space,
we observe the following superficial power counting rule
\bea
&\Omega = 4 - (n_x - 1) d - (2 - d/2) \, n_{\tau}  \nnb \\
& + v_{d} - 2 \, v_{\tau} - v_x \,.
\eea
Here $d$ means the dimension of the space-time. There is only one integral
in the coordinate space is taken as $4$, while others integrated out
are taken as $d$. Only positive $\Omega$ means divergence.

With this superficial divergence power counting rule,
the traces and logarithms in Eq. (\ref{divt}) are badly divergent.
However, the terms with $\Omega=4$ contribute to the vacuum and can be removed by a
global normalization factor in path integral, and the terms with $\Omega=2$ can be reduced
to logarithmic divergences in the dimensional regularization method.  Therefore there
are only logarithmic divergences which are physically meaningful.

Then by using the heat kernel method \cite{Avramidi:2000bm,Vassilevich:2003xt} directly,
we obtain the following divergence structures
from the contributions of $Tr\ln\Box$ in the Eq. (\ref{divt})
\bwt
\bea
{1 \over 2} {\bar \epsilon} Tr\ln\Box_{VV}&=&- {20 g^2 \over 3} H_1 -{(2 g^4 + G^4) \over 16} (v + \ch)^4 \,,
\eea
\bea
{1 \over 2} {\bar \epsilon} Tr\ln\Box_{\xi\xi}&=&
+ {g^2 \over 12} H_1 + {g^{'2} \over 12} H_2 
+ {1\over 12} {\cal L}_1 - {1 \over 24} {\cal L}_2
 - {1 \over 24} {\cal L}_3
-{1 \over 12} {\cal L}_4 - {1 \over 24} {\cal L}_5
\wsep + \frac{G^2 g^{'2} }{ 16} (v + \ch)^2 Z \cdot Z
 + \frac{g^2 (G^2 + g^2) }{ 32} (v+\ch)^2  W^+ \cdot W^-
\wsep - \frac{3}{4}\frac{(\partial^2 {h})^2}{(v + \ch)^2}
 + \frac{\partial^2 {h}}{4 (v + {h}) }  \left (2 g^2 W^+ \cdot W^- + G^2  Z \cdot Z \right ) 
-\frac{1}{8} (2 g^2 + G^2)  (v + \ch) \partial^2 {h}
\wsep -\frac{1}{64}(G^4 + 2 g^4) (v + \ch)^4 
 \,, \label{goldstone}
\eea
\bea
{1\over 2} {\bar \epsilon} Tr\ln\Box_{hh}&=&
-{1 \over 16} L_5
+{1 \over 32} \lambda [v^2 - 3 (v +\ch)^2 ] (G^2 Z\cdot Z + 2 g^2 W^+ \cdot W^-)
\wsep +{3 \over 32} \lambda^2 v^2 (v+\ch)^2 - { 9 \over 64} \lambda^2 (v+ \ch)^4
-{ 1 \over 16 } \lambda^2 v^4\,,
\eea
\bea
- {\bar \epsilon} Tr\ln\Box_{vu}= &=& -{2 g^2 \over 3} H_1 + {G^4 + 2 g^4 \over 32 } (v +\ch)^4\,,\eea\ewt
where ${16 \pi^2}/{\bar \epsilon}=(2/\epsilon - \gamma_E + \ln(4 \pi^2)$, $\gamma_E$ is
the Euler constant, and $\epsilon=4-d$.  For the convenience to construct counter terms,
we have deliberately add a sign in each term in the Eq. (\ref{divt}). In the Eq. (\ref{goldstone}),
we would like to mark terms proportional to $(\partial^2 {h})$, which
are impossible to be extracted if we work in the momentum space.

The divergence terms from the mixing terms with two
propagators are calculated by using the technique shown in \cite{Dutta:2004te}
and are given as
\bwt
\bea
-{\bar \epsilon \over 2} Tr(\rx_{\xi} \Box_{VV}^{-1} \lx_{\xi} \Box_{\xi\xi}^{-1})&=&
{g^2 g^{'2} \over 8} Z\cdot Z (v + \ch)^2
\nwsep - {g^2 + G^2 \over 8} (v+\ch)^2 (G^2 Z\cdot Z + 2 g^2 W^+ \cdot W^-)
\wsep - {1\over 2} (2 g^2 +G^2 ) \partial \ch \cdot \partial \ch\,,
\eea
\bea
- {\bar \epsilon \over 2} Tr(\rx_h \Box_{VV}^{-1} \lx_h \Box_{hh}^{-1})&=&
- {g^2 g^{'2} \over 8} Z\cdot Z (v + \ch)^2
\nwsep-{g^2 \over 8} (v + \ch)^2 (G^2 Z\cdot Z + 2 g^2 W^+ \cdot W^-)\,,
\eea
\bea
- {\bar \epsilon \over 2} Tr(X_{h\xi} \Box_{\xi\xi}^{-1} X_{\xi h} \Box_{hh}^{-1})&=&- {1\over 2} (t_{BB1} + t_{BB2} + t_{BC} + t_{CC})\,,
\eea
\bea
t_{BB1}&=&- {g^2 \over 6} H_1 - {g^{'2} \over 6} H_2 + {1\over 6} {\cal L}_1 + {1 \over 6} {\cal L}_2 + {1 \over 6} {\cal L}_3
+ {1 \over 6} {\cal L}_4 - {1 \over 6} {\cal L}_5
\wsep -{1\over 2} ({g^2 \over G^2 } -1)^2 {\cal L}_6 + {1\over 2} ({g^2 \over G^2 } -1)^2 {\cal L}_{10}
\nwsep -{ G^2 \over 4 } (\partial \cdot Z)^2 -{g^2 \over 2} (d \cdot W^{+}) (d \cdot W^{-})
\wsep - i {g^2 g^{'2} \over G} [(d \cdot W^{+}) (W^{-} \cdot Z) - (d \cdot W^{-}) (W^{+} \cdot Z)]\,,\label{ghmix}
\eea
\bea
t_{BB2}&=& - \frac{1}{4} {\cal L}_2   - \frac{1}{4} {\cal L}_3  - \frac{1}{2} {\cal L}_4   + \frac{1}{4} {\cal L}_5
\wsep - \frac{1}{16} (v + \ch)^2 (2 g^4 W^+ \cdot W^- + G^4 Z \cdot Z)
-\frac{3 \lambda }{16} (v + \ch)^2 (2 g^2 W^+ \cdot W^- + G^2 Z \cdot Z)
\wsep +\frac{\lambda}{16} v^2 (2 g^2 W^+ \cdot W^- + G^2 Z \cdot Z)
-\frac{1}{4 (v + \ch)} (2 g^2 W^+ \cdot W^- + G^2 Z \cdot Z) \partial^2 \ch
\,, \label{BB2}
\eea
\bea
t_{BC}&=&
({g^2 \over G^2 } -1)^2 {\cal L}_6 - ({g^2 \over G^2 } -1)^2 {\cal L}_{10}
\nwsep +{G^2 \over 2} (\partial \cdot Z)^2 + g^2 (d \cdot W^{+}) (d \cdot W^{-})
\wsep - i {g^2 g^{'2} \over G } [(d \cdot W^{+}) (W^{-} \cdot Z) - (d \cdot W^{-}) (W^{+} \cdot Z)]
 \,,
\eea
\bea
t_{CC}&=&
-{1\over 2} ({g^2 \over G^2 } -1)^2 {\cal L}_6 + {1\over 2} ({g^2 \over G^2 } -1)^2 {\cal L}_{10}
\nwsep - {G^2 \over 4} (\partial \cdot Z)^2 -{g^2 \over 2} (d \cdot W^{+}) (d \cdot W^{-})
\wsep + i {g^2 g^{'2} \over 2 G}[(d \cdot W^{+}) (W^{-} \cdot Z) - (d \cdot W^{-}) (W^{+} \cdot Z)]
\,.
\eea\ewt
Here the term proportional to $\partial^2 \ch$ in Eq. (\ref{BB2}) comes from the
$\sigma_{\xi\xi}$, and it is impossible to find such a compact divergence term if
we work in momentum space.

Due to the derivative coupling between the
Goldstone and Higgs boson, the term with four propagators (two Goldstone and two Higgs
propagators) also contributes divergences, which can be 
extracted out and given as 
\bwt
\bea
- {\bar \epsilon \over 4} Tr(X_{h\xi} \Box_{\xi\xi}^{-1} X_{\xi h} \Box_{hh}^{-1}X_{h\xi} \Box_{\xi\xi}^{-1} X_{\xi h} \Box_{hh}^{-1})
&=&- {1\over 12} {\cal L}_4 - {1 \over 24} {\cal L}_5\,.
\eea
\ewt
In the Weinberg's power counting rule, these divergences are counted as $p^4$ due to the
proper momentum assignment on the dimensionless couplings. While in Georgi-Manohar's power
counting rule, these divergences are counted as dimension $4$. In these
intermediate results, there are extra divergence structures as shown by those ${\cal L}_i$.
These extra intermediate divergence structures are defined as
\bea
{\cal L}_1&=&  \frac{g {g'}}{ 2} B_{\mu\nu}tr({\cal T} W^{\mu\nu})\cma
\eea
\bea
{\cal L}_2&=&i \frac{ g' }{ 2} B_{\mu\nu}tr({\cal T} [V^\mu,V^\nu])\cma
\eea
\bea
{\cal L}_3&=&i            tr( W_{\mu\nu}[V^\mu,V^\nu])\cma
\eea
\bea
{\cal L}_4&=&             [tr(V_\mu V_\nu)]^2\cma
\eea
\bea
{\cal L}_5&=&             [tr(V_\mu V^\mu)]^2\cma
\eea
\bea
{\cal L}_6&=&             tr(V_\mu V_\nu)tr({\cal T} V^\mu)tr({\cal T} V^\nu)\cma
\eea
\bea
{\cal L}_7&=&             tr(V_\mu V^\mu)[tr({\cal T} V^\nu)]^2\cma
\eea
\bea
{\cal L}_8&=&     \frac{g^2}{ 4} [tr({\cal T} W_{\mu\nu})]^2\cma
\eea
\bea
{\cal L}_9&=&  i \frac{g}{ 2} tr({\cal T} W_{\mu\nu})tr({\cal T} [V^\mu,V^\nu])\cma
\eea
\bea
{\cal L}_{10}&=& \frac{1}{ 2} [tr({\cal T} V_\mu)tr({\cal T} V_\nu)]^2\,.
\eea
These operators have been used to construct the set of complete operators of the EWCL when
C, P, T, and CP discrete symmetries are conserved \cite{Longhitano:1980iz,Longhitano:1980tm}.

To sum over all contributions yields the following total divergence structures as
\bwt
\bea
{\bar \epsilon} D_{tot} &=& - \frac{43 g^2}{6}  H_1 + \frac{g'^2}{6} H_2
- \frac{3}{16} {\cal L}_5 
-\frac{3}{32} (2 g^2 + G^2) (v +\ch)^2 (G^2 Z \cdot Z + 2 g^2 W^+ \cdot W^-)
\wsep + \frac{3}{8 (v +\ch)} \partial^2 \ch (G^2 Z \cdot Z + 2 g^2 W^+ \cdot W^-)
-\frac{3}{64} (2 g^4 + G^4 + 3 \lambda^2 ) (v +\ch)^4
\wsep - \frac{\lambda^2}{64} v^4 + \frac{3 \lambda^2}{32}  v^2 (v + \ch)^2
-\frac{1}{2} (2 g^2 + G^2)  \partial \ch \cdot \partial \ch
-\frac{1}{8} (2 g^2 + G^2) (v + \ch) \partial^2 \ch
-\frac{3 (\partial^2 \ch)^2 }{4 (v + \ch)^2}.
\label{prediv}
\eea
\ewt
It is remarkable that
terms related with the EOM of classic vector fields are exactly eliminated out, and
there is no trace of it at all.
However, here we observe there are some extra divergences, like ${\cal L}_5$, 
$\partial^2 \ch (G^2 Z\cdot Z + 2 g^2 W^+ \cdot W^-)$, and $(\partial^2 \ch)^2$, 
etc., which can not be eliminated. Terms proportional to $\partial^2 \ch$
are beyond the reach of the perturbation calculation method in the momentum space.  

If we stop here, then it seems
that our calculation procedure can not demonstrate the renormalizability of the
renormalizable theory, therefore it should be rejected as a
proper method to construct divergences of EWCL.
Fortunately, in the background field method, we can use the
EOM of classic fields.
By using the linear realized Higgs model as a guide and using the EOM of the Higgs field given
in Eq. (\ref{eomhiggs}) 
carefully, the total divergence structure can be reformulated as
\bwt
\bea
{\bar \epsilon} D_{tot} &=& -{ 43 \over 6} g^2 H_1 + {1 \over 6} g^{'2} H_2
\nwsep -{2 g^2 + G^2 \over 8} ( v+ \ch)^2 (G^2 Z \cdot Z + 2 g^2 W^+ \cdot W^-)
\wsep -{2 g^2 + G^2 \over 2} \partial \ch \cdot \partial \ch
\nwsep +{1 \over 32} (6 \lambda + 2 g^2 +G^2) \lambda v^2 (v+ \ch)^2
\wsep-{1 \over 64} [(6 g^4 + 3 G^4) + (2 g^2 + G^2) \lambda + 12 \lambda^2] (v+\ch)^4
\nwsep -{1 \over 16} \lambda^2 v^4\,.
\label{nowdiv}
\eea
\ewt
We observe that all the extra divergence 
structures ${\cal L}_i$ and $\partial^2 \ch$ are eliminated out.
This just indicates that the extra divergence structure just cancel out exactly
with each other, and even the terms like $(\partial Z)^2$ do not appear in the
total divergence structures. This is exactly the meaning of the
renormalizability. Even no extra gauge fixing term of the background
fields should be added to the Lagrangian, and
the equations of motion of the background fields
are sufficient. This is a pleasant result and really support our
renormalization procedure.

By using the reverse Stueckelberg transformation \cite{Grosse-Knetter:1992nn}, the background Goldstone
fields (the longitudinal component of the vector bosons) 
can be restored in the Lagrangian. Therefore the Lagrangian contains
the correct dynamic degrees of freedom at low energy region.
It is interesting to observe that these divergence terms (counter terms) are counted as
$p^4$ in the Weinberg's power counting rule and simply dimension $4$ operators
in the Georgi-Manohar's power counting rule. In Weinberg's power counting rule,
these $p^4$ divergences can be absorbed by refining operator's renormalization constants,
which induce the anomalous dimension matrix among operators;
while in the Georgi-Manohar's power counting rule, these divergences
of dimension $4$ operators can be absorbed by redefining the renormalization constants of
fields, couplings and mass parameter, like the standard renormalization procedure.

Here we would like to point out that in order to justify the perturbation method, in the
realistic consideration, even we can define both these two power counting rules in our
theory, we must require that all couplings in the loop expansion should 
be weak couplings in order to guarantee the validity of the perturbation expansion.

There are terms which contribute to the vacuum. After using the normalization of the
partition functional, these terms can be eliminated out. 

We would like to mention that the coefficients of $H_1$ and $H_2$
have the correct value which contribute
to the $\beta$ functions of the gauge couplings $g$ and $g'$, as calculated 
with a linearly realized Higgs doublet, respectively. According to the well known
results \cite{Gross:1973id,Politzer:1973fx,'tHooft:1973us}, the $\beta$ function of non-Abelian gauge theory can be expressed as
\bea
\frac{g^3}{16 \pi^2} \left( -\frac{11}{3}\times 2  + \frac{1}{3} \times \frac{1}{2} \right)\,.
\eea
While the $\beta$ function of Abelian gauge theory can be expressed as
\bea
\frac{{g'}^3}{16 \pi^2} \left( \frac{1}{3} \times \frac{1}{2} \right)\,.
\eea
Our results given in Eq. (\ref{nowdiv}) agree with these well-known results.

One interesting observation is that the Goldstone particles only contribute half of
quantities of a Higgs doublet. This can be read out from Eq. (\ref{goldstone}).  In the
nonlinear realized Higgs, another half comes from the Goldstone-Higgs mixing contribution,
as shown in Eq. (\ref{ghmix}). 
To produce these correct $\beta$ functions of gauge couplings 
also convince us that our renormalization procedure 
can be correct for the systematic renormalization of the EWCL.

Here we observe that the coefficients of the terms
$( v+ \ch)^2 (G^2 Z \cdot Z + 2 g^2 W^+ \cdot W^-)/8$
and $\partial \ch \cdot \partial \ch/2$ are equal, and this is not accidental. If
we reformulate the Lagrangian in its linear form,
the combination of these two terms just yields the term
$(D\phi)^{\dagger}\cdot(D \phi)$.  We have used this as a guideline to eliminate the
ambiguity in the usage of the EOM of classic Higgs field.

\section{Divergences of $O(p^2)$ without a Higgs boson}
When the Higgs field is assumed to be decoupled from low energy
phenomenology and only $O(p^2)$ 
operators (in Weinberg's power counting rule) are considered,
the classic Lagrangian is modified as
\bea
\label{hml}
{\cal L} &=& - H_1 - H_2 + {v^2 \over 4} tr[V \cdot V] \,. 
\eea
After taking into account the quantization in the background field 
method, the active degree of freedom
of the system includes quantum vector boson, quantum Goldstone boson, and
ghost fields. By working in the mass eigenstates, the gauge
fixing terms are determined as
\bea
{\cal L}_{GF,A}&=&-{1 \over 2} (\partial\cdot\hA  \wsep - i e (\hWm \cdot \bWp - \hWp \cdot \bWm))^2\cma
\eea
\bea
{\cal L}_{GF,Z}&=&-{1 \over 2} (\partial\cdot\hZ - {1\over 2} G v  \xi_Z
\wsep + i {g^2 \over G} (\hWm \cdot \bWp - \hWp \cdot \bWm))^2\cma
\eea
\bea
{\cal L}_{GF,W}&=&- (d\cdot\hWp + {1\over 2} g v \xi_W^+
      + i {g^2 \over G} \bZ \cdot \hWp \wsep - i {g^2 \over G} \bWp \cdot \hZ
	+ i e \bWp \cdot \hA)\nnb\\&&(d\cdot\hWm + {1\over 2} g v \xi_W^-
      - i {g^2 \over G} \bZ \cdot \hWm \wsep + i {g^2 \over G} \bWm \cdot \hZ
	- i e \bWm \cdot \hA)\,.
\eea
Compared with the case with a nonlinearly realized Higgs boson,
these gauge fixing terms are obtained from Eqs. (\ref{gfx1}---\ref{gfx2}) by setting $\ch=0$. 
These gauge fixing terms are covariant background field
gauges when we rotate they back to weak interaction eigenstate
basis. The bilinear terms of the quantum fluctuations can be
expressed as
\bea
{\cal L}_{quad}&=&- \bbbkf{ {1\over 2 } {\widehat V_{\mu}^{\dagger a}} \Box^{\mu\nu,ab}_{V\,V} {\widehat V_{\nu}^b}
+ {1\over 2} \xi^{\dagger i}  \Box_{\xi\,\xi}^{ij} \xi^j
\wsep + v^a  \Box_{vu}^{ab} u^b
\wsep
+ {1\over 2} {\widehat V_{\mu}^{\dagger,a}} {\stackrel{\leftharpoonup}{X}}_{\xi}^{\mu,aj} \xi^j
+ {1\over 2} \xi^{\dagger,i} {\stackrel{\rightharpoonup}{X}}_{\xi}^{\nu,ib} {\widehat V_{\nu}^b} }\,.
\eea
These operators can be obtained from Eq. (\ref{lag}) by setting $\ch=0$. While other parts, like
the gauge potential, etc, are not modified.
After performing the path integral to integrate out all quantum fluctuations, the 
one-loop effective Lagrangian can be expressed as
\bea
\int_x {\cal L}_{1-loop}&=& Tr\Box_{\bar{c}c}  - {1\over 2} \mbra{ Tr\ln\Box_{VV}
+Tr\ln\Box_{\xi\xi}  
\wsep - Tr(\rx_{\xi} \Box_{VV}^{-1} \lx_{\xi} \Box_{\xi\xi}^{-1})
 + \cdots}
\,.
\label{divtop2}
\eea
After extracting divergences, we obtained the following divergence structure
\bea
{\bar \epsilon} D_{tot} &=&-\frac{29 }{4} g^2 H_1 + \frac{1}{12} {g'}^2 H_2 
\wsep + \frac{1}{12} {\cal L}_1 -  \frac{1}{24} {\cal L}_2
-  \frac{1}{24} {\cal L}_3 -  \frac{1}{12} {\cal L}_4 -  \frac{1}{24} {\cal L}_5
\wsep + \frac{3 (G^2 + g^2) }{8}   \times {v^2  \over 4} tr[V \cdot V]
\wsep - \frac{3 {g'}^2}{8}  \times {v^2  \over 4} (Tr[{\cal T} V])^2
\wsep -\frac{3}{64} (2 g^4 + G^4) v^4\,.
\eea
We observe that without the help of the Higgs field, those
divergences can not be eliminated out any more. 
Here we mark two facts: 1) The $\beta$ functions of the
gauge couplings are correct when we count the active degree of freedom.
The $\beta$ function of $g$ is given as
\bea
\beta_g&=&\frac{g^3}{16 \pi^2} \left ( -\frac{11}{3} \times 2 + \frac{1}{6} \times \frac{1}{2} \right )\,.
\eea
The $\beta$ function of $g'$ is given as
\bea
\beta_{g'}&=& \frac{{g'}^3}{16 \pi^2} \left (\frac{1}{6} \times \frac{1}{2}  \right )\,.
\eea
The $\beta$ function of the $v^2$ is given as
\bea
\beta_{v^2}=\frac{v^2}{8 \pi^2} \times (\frac{3 (G^2 +g^2) }{8})\,.
\eea
2) The $\beta$ functions for the anomalous couplings are completely determined
by the evaluation of the exact Goldstone propagator, which can be formulated as
\bea
\beta_{\beta} &=& \frac{1}{8 \pi^2} \times ( - \frac{3 {g'}^2}{8} )\,,\\
\beta_{\al_1} &=& \frac{1}{8 \pi^2} \times \frac{1}{12}\,,\\
\beta_{\al_2} &=& \frac{1}{8 \pi^2} \times ( - \frac{1}{24} )\,,\\
\beta_{\al_3} &=& \frac{1}{8 \pi^2} \times ( - \frac{1}{24} )\,,\\
\beta_{\al_4} &=& \frac{1}{8 \pi^2} \times ( - \frac{1}{12} )\,,\\
\beta_{\al_5} &=& \frac{1}{8 \pi^2} \times ( - \frac{1}{24} )\,.
\eea
By solving these RGE of the anomalous couplings, the logarithmic contributions
of the heavy Higgs can be found and reproduce the
"screening" effects of Higgs \cite{Appelquist:1980ix,Veltman:1976rt} to low energy phenomenology.

Due to our definition of the 
covariant differential operator given in Eq. (\ref{covdo}) and Euclidean space, we observe
several sign differences in our divergences when compared with the
well-known results \cite{Longhitano:1980iz,Longhitano:1980tm,Herrero:1993nc,Herrero:1994iu,Dittmaier:1995ee}. 
Compared with the well-known paper given by Appelquest and Wu \cite{Appelquist:1993ka}, 
our $\beta$ parameter has a sign difference due to the Euclidean space, and 
our triple anomalous couplings have the opposite signs due to the definition of the
covariant differential operator given in \cite{Appelquist:1993ka} has the following form
\bea
V_{\mu}= U^\dagger \partial_{\mu} U + i U^\dagger  W_{\mu} U - i B_{\mu}\,.
\eea
After realizing these differences, we find complete agreements 
between our results and with these well-known results.

As matter of fact, 
sign differences existed in literatures are due to the fact there exist discrete symmetries
in the definition of the partition functional
\bea
g\rightarrow \pm g \,,\,\, g' \rightarrow \pm g'\,,\,\, v\rightarrow \pm v\,,
\eea
which can explain why there exist arbitrariness in the definition of
the covariant differential operator. In the path integral,
vector boson and Goldstone both are assumed to be real fields,
there exists another type of discrete symmetries in the field 
definition which corresponds to the above
discrete symmetries in couplings (here we regard $v$ as a coupling)
\bea
W \rightarrow \mp W\,,\,\, B \rightarrow \mp B\,,\,\, \xi \rightarrow  \mp \xi\,.
\eea
These symmetries cause convention problems in the anomalous couplings at $O(p^4)$ order.

The fact that our renormalization procedure can reproduce the 
divergences of these $O(p^2)$ operators
reinforces our belief that our renormalization procedure is correct.

\section{BRST transformations}
The BRST global transformation \cite{Becchi:1975nq} is very 
important to formal proof of the
renormalizability of the spontaneous symmetry 
breaking models and gauge independence
of the S-matrix. Here we
ask the question: in our calculation procedure 
with nonlinear realization, 
is there such a global transformation?
The answer is yes. Below we explicitly construct 
the BRST transformation in the BFM formalism
in the nonlinearly realized Higgs model and in 
$O(p^2)$. We have referred  the BRST transformation without spontaneous
symmetry breaking \cite{Kluberg-Stern:1974xv} and in
the linearly realized Higgs doublet \cite{Denner:1994xt}
for our construction. 

We start with the BRST transformation with a linearly realized
Higgs field. Due to the fact that in the background field method we can
choose different gauges for the classic and quantum fields,
the background Goldstone fields can be eliminated from the
Lagrangian by choosing the unitary gauge for the background
vector fields. So the Higgs field can be
parameterized as
\bea
\Phi &=& \frac{1}{\sqrt{2}} \left ( {\bf 1} (v + \rho + \qrho) + {\bf i} {\hat \phi}^i T^i \right )\,.
\eea

The gauge fixing terms in this linearly realized Higgs
field can be expressed as
\bea
F_Y &=& \partial \cdot {\hat B} + {g'} (v + \rho) \phi^3 \,,\nnb\\
F^i_W &=& (\de^{ij} \partial + g f^{ijk} W^{j}) 
\cdot {\hat W}^{k}  + g (v + \rho)  \phi^i\,.
\eea
Here we notice that in these gauge fixing terms the $SU_c(2)$ isospin
symmetry originated from $SU(2)$ 
has been explicitly broken by $F_Y$, so we can not
rotate $\phi^1$, $\phi^2$, and $\phi^3(=\phi^Z)$ any more. 
Here we observe that the quantum fluctuation of Higgs field, $\qrho$, 
does not enter into the gauge fixing terms. 

For both the renormalizable linearly realized Higgs model
and the EWCL up to $O(p^2)$ operators,
there exists a universal rotation matrix for all
the gauge group parameters, quantum vector fields, ghost fields, which can be
expressed as
\begin{displaymath}
\left (
\begin{array}{cccc}
 c_w & s_w & 0 &0 \\
 s_w & -c_w & 0 &0 \\
0 & 0 & \frac{1}{\sqrt{2}} &  \frac{{\bf i}}{\sqrt{2}} \\
0 & 0 & \frac{1}{\sqrt{2}} & -\frac{{\bf i}}{\sqrt{2}}
\end{array}
\right )\,.
\end{displaymath}
Then we observe that this rotation changes 
the functional measure with a trivial constant number, which 
is innocent and does not
affect physics. After this rotation, the
index of vector fields $a$ changes from $\{Y, 3, 1, 2\}$ to $\{A, Z, W^+, W^-\}$. While there
are only three Goldstone, their index can be represented
as $i$, which is changed from $\{3(Z), 1, 2\}$ to $\{Z, W^+, W^-\}$,
due to the remnant $U_{\textrm{EM}}(1)$ symmetry.
While the group structure constants $f^{ijk}$ of $SU(2)$ also
are also changed and the new group structure
constants $g^{ijk}$ with indices as $i=\{Z, W^+, W^-\}$ are determined by the commutation relations
of generators, $T^Z$, $T^+$, and $T^-$.
\bea
[T^+, T^-]=T^3\,,\,\,[T^3, T^+] = T^+\,,\,\,[T^-, T^3] =T^-\,.
\eea
We have $g^{+-3}=g^{3++}=g^{-3-}=-g^{-+3}=-g^{+3+}=-g^{3--}=1$.

The BRST transformation of the quantum system  in the mass eigenstate
basis can be formulated as
\bea
\de_B {\hat V}^a_{\mu} &=& - D^{ab}_{\mu}({\overline V} + {\hat V}) u^b\,,\nnb\\
\de_B u^a &=& -{ \bf i \over 2} \eta^{ai}  g^{ijk} \omega^{j b} \omega^{k c} u^b u^c\,,\nnb\\
\de_B (\phi^\dgr)^i &=& - (v + \crho + \qrho ) (u^{\dgr})^a \eta^{ai}   \,,\nnb\\
\de_B {\qrho} &=& \frac{1}{4} (u^{\dgr})^a \eta^{ai}  \phi^i \,,\nnb\\
\de_B v^a &=& B^a\,,\nnb \\
\de_B B^a &=& 0\,.
\label{lBRST}
\eea
The charge conjugate has been added to guarantee the correct
charge number for each equations.
The $i$ in the $\de_B u^a$ indicates the
change of from real fields to complex fields.

The $\eta$ is a $4 \times 3$ matrix, which is given as
\begin{displaymath}
\left (
\begin{array}{ccc}
e  & 0 & 0 \\
- \frac{g^2}{G}  & 0 & 0 \\
0 & g & 0 \\
0 & 0 & g
\end{array}
\right )\,,
\end{displaymath}
and $\omega$ is a $3 \times 4$ matrix, which is given as
\begin{displaymath}
\left (
\begin{array}{cccc}
s_w  & -c_w & 0 & 0\\
0 & 0 & 1 & 0\\
0 & 0 & 0 & 1
\end{array}
\right )\,.
\end{displaymath}

In order to guarantee the derivative power counting
rule, the BRST transformation 
$\de_B$ is assumed to change the power by the unit $+1$,
and ghost fields are assumed to be $[v]_p=[u]_p=0$.
The matrix $\eta$ contains the gauge couplings and
$[\eta]_p=1$ which
guarantees that this power counting rule can hold.
While for the mass dimension power counting rule,
$\de_B$ is also assumed to change the power by $+1$,
and ghost fields are assumed to be $[v]_m=[u]_m=1$.
The matrix $\eta$ is dimensionless in mass dimension
power counting rule.

In both these two power counting rule, the matrix $\omega$ is
dimensionless.

However, both these two assignments can not be 
consistent for the last equation.
Anyway, after integrating out the auxiliary field $B^a$ and
using the on-shell condition for the $u$ type ghost fields
given as
\bea
\frac{\de F^a}{\de \al^{\bar b} } u^{\bar b} &=& 0\,,
\eea
the Lagrangian enjoys the on-shell BRST transformation
and can be consistent with both
these two power counting rules.

As shown in \cite{Becchi:1975nq}, the ghost-BRST is sufficient for the
demonstration on the perturbation renormalizability 
and gauge independence of S-matrix.

The relation between the linearly realized and nonlinearly
realized Higgs and Goldstone fields are given as \cite{Grosse-Knetter:1992nn}
\bea
(v + \crho + \qrho) + \phi^i T^i &=& (v + h + \qh) U^{i \zeta^i T^i}\,,
\eea
with the relations between the component fields given as
\bea
\crho&=& (v + h) \cos \zeta -v \,,\\
\qrho &=& \qh \cos\zeta \,,\\
\phi^i &=& (v + h+ \qh) {\hat \zeta}^i \sin \zeta \,.
\eea
Where $\zeta=\sqrt{\zeta^{i,2}}$, and ${\hat \zeta}^i=\zeta^i/\zeta$.
The $\zeta^i$ are the phase angle which parameterize the Goldstone
fields with $\zeta^i=2 \xi/(v + h)$.

By checking the gauge transformation of 
Higgs fields, the BRST transformation of quantum
fluctuations in mass eigenstate basis is modified as
\bea
\de_B {\hat V}^a_{\mu} &=& - D^{ab}_{\mu}({\overline V} + {\hat V}) u^b\,,\nnb\\
\de_B u^a &=& -{ \bf i \over 2} \eta^{ai}  g^{ijk} \omega^{j b} \omega^{k c} u^b u^c\,,\nnb\\
\de_B (\xi^\dgr)^i &=& - {1 \over 2} (v + h) (u^{\dgr})^a \eta^{ai}   \,,\nnb\\
\de_B {\qh} &=& 0 \,,\nnb\\
\de_B v^a &=& B^a\,,\nnb \\
\de_B B^a &=& 0\,.
\label{nlBRST}
\eea
The reason for
the $\de_B {\qh}=0$ can be easily understood: 
in the nonlinear form, $\qh$ field is a $SU(2)$ singlet.
But the $\de_B {\qh}$ poses a new dilemma
for both power counting rules if we want to assign powers
to BRST transformation, compared with the case 
in the linear form given in Eq. (\ref{lBRST}),
where both these two power counting rules sustain with on-shell
ghost fields.
However, the Lagrangian with a nonlinearly realized
Higgs fields is consistent with the relation $\de_B {\qh}=0$.

When the Higgs scalar decouples or there is no Higgs at all, the 
BRST transformation of quantum
fluctuations in mass eigenstate basis in the EWCL up to $O(p^2)$ is modified as
\bea
\de_B {\hat V}^a_{\mu} &=& - D^{ab}_{\mu}({\overline V} + {\hat V}) u^b\,,\nnb\\
\de_B u^a &=& -{ \bf i \over 2} \eta^{ai}  g^{ijk} \omega^{j b} \omega^{k c} u^b u^c\,,\nnb\\
\de_B (\xi^\dgr)^i &=& - {1 \over 2} v \, (u^{\dgr})^a \eta^{ai}   \,,\nnb\\
\de_B v^a &=& B^a\,,\nnb \\
\de_B B^a &=& 0\,.
\eea

These constructions just show that
there is no difficulty to construct the BRST transformation even with nonlinear
realization.  However, the BRST transformation does not necessarily
guarantee the renormalizability of the nonlinear realization of the spontaneous
symmetry breaking for any higher loop order. Even 
in the nonlinearly realized Higgs boson,
the Feynman rules of Goldstone bosons and the
superficial divergence power counting rule make the renormalizability of the
model obscure.

\section{Discussions and Conclusions}
In this paper, by using the dimensional regularization (a symmetry preserving regularization
scheme) and Feynman-'t Hooft gauge in the background field method,
we show how our calculation procedure can demonstrate the
renormalizability of the nonlinearly realized Higgs in the SM,
can produce correct $\beta$ functions of gauge couplings, and can reproduce 
well-known results on the divergences of the $O(p^2)$ operators in the EWCL.
Compared with
those calculations with Landau gauge \cite{Longhitano:1980iz,Longhitano:1980tm,Herrero:1993nc,Herrero:1994iu,Dittmaier:1995ee},
our results show that those divergences are gauge independent
if a renormalizable gauge are taken.
In the $R_\xi$  gauge, the ordering of performing loop integrals and
taking the limit $\xi \rightarrow \infty$ can not be exchanged. 
This may explain why the divergences extracted by using the unitary gauge
can not agree with other renormalizable gauges.
We conclude that our method can serve as a reliable method to extract the
divergences of the EWCL.  

Below we would like remark on the regulation method existed in literatures.
Reference \cite{Hagiwara:1992eh,Hagiwara:1993ck} introduces a 
Higgs field as a regulator, after renormalization 
the effective Lagrangian the Higgs was taken to the 
decoupling limit. This procedure can not
produce the correct $\beta$ functions of gauge 
couplings and anomalous couplings of the EWCL
without a Higgs, since the basic degree of freedoms are different before and after
the decoupling limit.  A careful matching must be made 
between the EWCL with a Higgs and the EWCL 
without a Higgs. Only after this matching procedure,
the $\beta$ functions for anomalous couplings can be correctly obtained.

In the Slavnov's scheme\cite{slavnov}, higher dimensional
covariant operator are introduced as a regulator. Such a regularization
scheme has been used by 
J. J. van der Bij and B. Kastening \cite{vanderBij:1997ec} to study the
radiative correction of two point functions in the EWCL.
By using the standard superficial 
divergence counting rule, it is obviously that this
regularization method can improve the divergence power counting
behavior but can not lead to a consistent treatment 
to all divergences in the theory \cite{Martin:1994cg}.
The biggest trouble is that it can not produce the correct $\beta$ function for
the gauge couplings in the renormalizable gauge theory, 
as pointed out in \cite{Martin:1994cg,Leon:1995nm}, which also means
the breaking of unitarity of the S-matrix. 
For us, to use it as a regulator to nonlinear gauged sigma model 
seems to be more problematic.

Recently, Y.L. Wu has proposed  \cite{Wu:2003dd} a new symmetry preserving regularization
scheme. Whether this scheme can be used to perform systematic renormalization of
low energy QCD chiral Lagrangian or EWCL needs further study.

About our renormalization procedure, there are two remarks on it:

1) Although we have explicitly demonstrate the
renormalizability at one-loop level,
and have constructed the BRST transformation formally,
it is reasonable to ask whether the renormalizability 
in this calculation is accidental or necessary, 
like the one-loop renormalizability for the gravity?
Can it work at two-loop level or higher? 
These questions are worthy of our future study.

2) Due to the semi-simple structure
of the $SU(2) \times U(1)$ symmetry, and the symmetry breaking
pattern $SU(2) \times U(1) \rightarrow U(1)$, in the vector
sector and ghost sector, we can have both the weak interaction eigenstates and
mass eigenstates. It is natural to ask whether our 
calculation  procedure in the weak interaction eigenstates are also viable.
In weak interaction eigenstate basis, the BRST of $SU(2)\times U(1)$ can be more
easily constructed.  In principle, although massless vector
fields theory has the problem of infrared divergences, when
we only consider the ultraviolet divergences, it is possible to
calculate in the weak interaction eigenstate basis.
The computation in the weak interaction eigenstates is 
worthy of examination in our future work.

By using these concepts and methods,
we will explore the systematic renormalization
of the EWCL in our future works \cite{preparation}.

\acknowledgments{
The author would like to
thank Prof. M. Tanabashi for helpful and stimulating discussions.
The author also would like to thank Prof. K. Hagiwara
for his constant encouragement during this project.
The work is supported by the JSPS fellowship program. 
}

\end{document}